\documentclass[conference]{IEEEtran}
\IEEEoverridecommandlockouts
\usepackage{cite}
\usepackage{amsmath,amssymb,amsfonts}
\usepackage{algorithmic}
\usepackage{graphicx}
\usepackage{textcomp}
\usepackage{xcolor}
\def\BibTeX{{\rm B\kern-.05em{\sc i\kern-.025em b}\kern-.08em
    T\kern-.1667em\lower.7ex\hbox{E}\kern-.125emX}}
\begin{document}

\title{Adaptive Robotic Arm Control with a Spiking Recurrent Neural Network on a Digital Accelerator
\thanks{This research was partially supported by the Spanish grant MINDROB (PID2019-105556GB-C33/AEI/10.13039/501100011033) and by the SMALL (PCI2019-111841-2/AEI/10.1309/501100011033) EU CHIST-ERA.}
}

\author{\IEEEauthorblockN{Alejandro Linares-Barranco$^1$, Luciano Prono$^2$, Robert Lengenstein$^3$, Giacomo Indiveri$^4$, Charlotte Frenkel$^5$}
\IEEEauthorblockA{$^1$\textit{Robotics and Computer's Tech Lab}. \textit{University of Sevilla}. Sevilla, Spain. (alinares@atc.us.es)\\
$^2$\textit{Dpto. of Electronics and Telecom.} \textit{Politecnico di Torino}. Torino, Italy\\
$^3$\textit{Institute of Theoretical Computer Science.} \textit{Graz University of Technology}. Graz, Austria\\
$^4$\textit{Institute of Neuroinformatics.} \textit{University of Zurich}. Zurich, Switzerland\\
$^5$\textit{Dept of Microelectronics.} \textit{Delft University of Technology}. Delft, The Netherlands}
}

\maketitle

\begin{abstract}
With the rise of artificial intelligence, neural network simulations of biological neuron models are being explored to reduce the footprint of learning and inference in resource-constrained task scenarios. A mainstream type of such networks are spiking neural networks (SNNs) based on simplified \textit{Integrate and Fire} models for which several hardware accelerators have emerged. Among them, the ``ReckOn'' chip was introduced as a recurrent SNN allowing for both online training and execution of tasks based on arbitrary sensory modalities, demonstrated for vision, audition, and navigation. As a fully digital and open-source chip, we adapted ReckOn to be implemented on a Xilinx Multiprocessor System on Chip system (MPSoC), facilitating its deployment in embedded systems and increasing the setup flexibility. We present an overview of the system, and a Python framework to use it on a Pynq ZU platform. We validate the architecture and implementation in the new scenario of robotic arm control, and show how the simulated accuracy is preserved with a peak performance of 3.8M events processed per second.
\end{abstract}


\section{Introduction}

Spiking neural networks (SNNs) are a particular variant of neural networks, which use pulse-based signals to process information. Unlike traditional artificial neural networks (ANNs), which compute the weighted sum of inputs through multiply-accumulate operations~\cite{Rabinovich_etal06}, SNNs rely on accumulation operations triggered by binary spikes~\cite{MAASS19971659}. This property allows for energy efficiency advantages in sparse scenarios~\cite{Diehl_Neftci2016_SNNlowpower}.
The implementation of spiking neural network accelerators in hardware aims to exploit this efficiency advantage through highly parallel architectures. It is a research area that has attracted much attention in recent years and utilized both analog and digital design techniques, both synchronous and asynchronous, based on ASICs \cite{spinnaker_ijcnn2008,Sawada_etal16,Moradi_etal18,Benjamin_etal14,Davies_etal18,reckon_frenkel_2022} or FPGAs \cite{Cheung_Gunther_SNNAcc_2012,Khodamoradi_Kastner_S2N2_2021}.
The state of art has shown that these SNN accelerators have great potential for use in a wide range of applications, such as speech and gesture recognition, or robot navigation and control~\cite{reckon_frenkel_2022,Milde_etal17b,Donati_etal19}. These applications require real-time data processing and high energy efficiency, making SNN accelerators an ideal solution.

\begin{figure}[t]
\centering
 \includegraphics[width=0.45\textwidth]{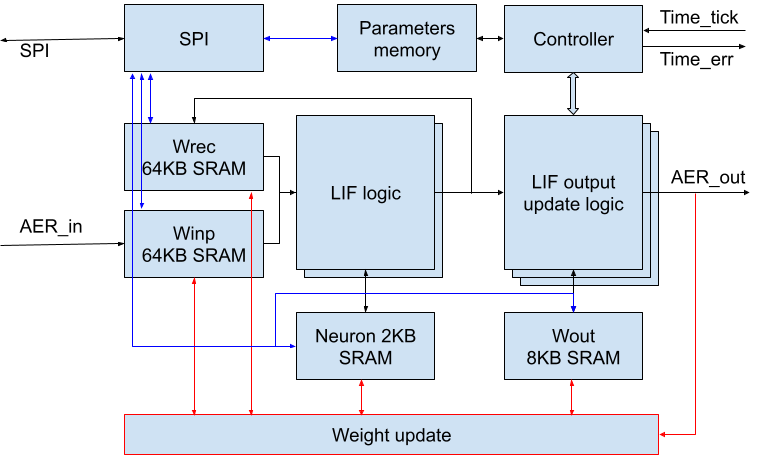}
 \caption{Block diagram of the ReckOn accelerator (simplified from \cite{reckon_frenkel_2022}).}
  \label{fig:diagramReckonSimplified}
\end{figure}

\begin{figure*}[h]
\centering
 \includegraphics[scale=0.37]{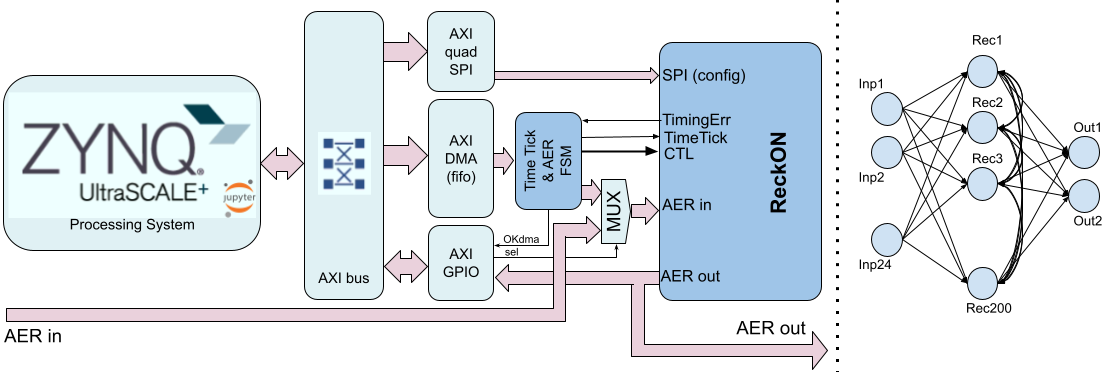}
 \caption{Left: Block diagram of ReckOn on the Pynq-ZU board. Dark blue blocks are implemented in the PL to support ReckON, while light blue blocks correspond to the PS and IPs from Xilinx library for connectivity. Right: RSNN used in the experiments.}
  \label{fig:diagramabloques}
\end{figure*}

In this paper, we demonstrate an SNN used to perform online adaptive robotic arm control using an open-source recurrent SNN accelerator denoted as ``ReckOn''~\cite{reckon_frenkel_2022}, which is implemented in the Verilog hardware description language. Our contributions are two-fold: (i)~we integrate ReckOn as part of a full system deployed on a Xilinx multiprocessor programmable system (MPSoC), the Zynq UltraScale+, to seamlessly configure and interact with the hardware accelerator through an online Python interface running on an embedded Jupyter server; (ii)~using a Pynq-ZU board, we demonstrate the full system in the real-world task scenario of adaptive robotic arm control, exploiting the ability of ReckOn to learn online.

This paper is organized as follows. Section~\ref{sec:reckon} first describes the ReckOn accelerator in brief. Section~\ref{sec:implementation} then explains the details and additional circuits needed for the system integration. Finally, Section~\ref{sec:results} presents the results, after which we summarize the main outcomes.

\section{The RSNN accelerator ReckOn} \label{sec:reckon}
The ability to learn short- and long-term temporal dependencies in embedded hardware is one of the key missing elements needed to improve the robustness of autonomous devices in the real-world, guarantee user privacy, and reduce reliance on the cloud~\cite{reckon_frenkel_2022}. To address this challenge, ReckOn proposes an RSNN processor that enables supervised learning over seconds while keeping a millisecond-range temporal resolution. As the vanilla backpropagation-through-time (BPTT) training algorithm requires backpropagating error information through the network dynamics, its memory requirements do not fit resource-constrained autonomous devices for learning over long timescales. To solve this challenge, Bellec et al.~proposed the eligibility propagation (e-prop) algorithm~\cite{bellec2020solution}, which is a forward-mode learning algorithm based on eligibility traces (ETs) that provides a bio-plausible alternative that approximates BPTT. ReckOn implements a modified version of e-prop that is fully local in both space and time, thereby drastically reducing memory requirements by only scaling with the number of neurons, instead of the number of synapses~\cite{reckon_frenkel_2022}.


Figure \ref{fig:diagramReckonSimplified} shows a simplified diagram of the ReckOn accelerator architecture. The RSNN architecture is based on a layer of 256 recurrently connected leaky integrate-and-fire (LIF) neurons, to which 256 input neurons and from which 16 output readout neurons are fully connected. The corresponding recurrent, input, and output weight matrices ($W_{rec}$, $W_{inp}$, $W_{out}$), containing 8-bit weights, as well as the neuron state ($Neuron SRAM$) are stored in on-chip SRAM.

Typically, temporal multiplexing and asynchronous communication techniques are used to transmit the spikes from source neurons to destination ones from one chip to another. The most common one is the Address-Event-Representation (AER)~\cite{AER}. 
ReckOn has a spiking AER input bus to allow interfacing with arbitrary neuromorphic sensors, such as the retina \cite{Serrano13} and cochlea \cite{Jimenez17}, and an SPI bus for initial configuration and debugging. The optional $Time\_tick$ signal allows synchronizing the internal neural processing steps with an external timer, which indicates the temporal resolution of the input data. The module ($Weight\_update$) carries out updates based on the modified e-prop algorithm. ReckOn also exploits sparsity on input data and weight updates to reduce the energy footprint. Prototyped in a 28-nm CMOS node, ReckOn was demonstrated for real-time task-agnostic learning of gesture recognition, keyword spotting, and navigation tasks within power budgets not exceeding 50\,$\mu$W~\cite{reckon_frenkel_2022}. Implemented in the Verilog hardware description language, it is available in open source~\cite{Frenkel_ReckOn_git}.

\section{MPSoC implementation}
\label{sec:implementation}


We deployed ReckOn on a Xilinx MPSoC, model Zynq Ultrascale+ XCZU5EG, for the Pynq-ZU board \cite{pynq-zu-git}. This chip from Xilinx consists of an embedded processing system (PS) based on four 64-bit ARM Cortex A53 processors, two ARM Cortex R5F real-time processors, an ARM MALI 400MP GPU-type graphics processor, along with an FPGA (programmable logic, PL) on which to deploy the desired circuitry and communicate with the PS. Based on a Petalinux operating system, we deploy a Jupyter Notebook server in the PS, which allows driving the FPGA using Python through Pynq libraries. We use it to program the PL, to preprocess data to streamline the PS-PL communication, and to collect and analyze the output from ReckOn. The Xilinx Vivado tool, version 2021.2, has been used for development. The Jupyter Notebook server running on the Pynq-ZU board can be operated from a computer, through USB connection and also allows uploading to the PL the bitstream files generated from Vivado.


Fig.~\ref{fig:diagramabloques} (left panel) shows the integration of ReckOn in the Xilinx MPSoc system, resulting from five key design decisions, as follows.
\begin{itemize}
    \item ReckOn's input AER interface can be chosen to come either (i) from an actual AER sensor interfaced with the board, or (ii) from the PS through an AXI-DMA-interfaced FIFO storing tuples of address-events and timestamps (AE,TS) of samples in a given dataset. This mux-selectable scheme improves flexibility, facilitates debugging and testing of the accelerator, and is easier to deploy with offline datasets.
    \item To ensure precise timing control within a sample, we use two concatenated finite state machines ($Time\_tick~\&~AER~FSM$ block) instead of an AXI-GPIO interface, whose latency would lead to timing distortion. One FSM keeps reading the FIFO containing the events of a dataset sample, while the second controls the sending of these events to ReckOn while ensuring correct timing. More specifically, these FSMs are driven by the PS based on the dataset requirements and take care of the start and end of each sample, of the Req/Ack handshake for each event in a given sample, of providing ticks defining the time reference, of enabling the weight update module for learning with training samples (disabling learning for inference with test samples), and of reading back Reckon's output to provide it to the PS.
    \item The output of ReckOn, provided at each timestep (regression) or at the end of a sample (classification), also follows an AER format. It can either be sent to the PS through an AXI-GPIO interface, or forwarded to the outside world via the Pynq-ZU board.
    \item ReckOn requires a configuration interface based on the SPI bus to configure the network, the network weights, and the different configuration parameters available. It is interfaced via a Xilinx AXI-quad-SPI interface that is being used in basic SPI mode.
    \item As ReckOn is implemented in the PL, the SRAM memories of the original design have been swapped for block RAM (BRAM) resources available inside the FPGA.
\end{itemize}  

\section{Experimental Results}
\label{sec:results}

\begin{figure}[t]
 \centering
 \includegraphics[scale=0.35]{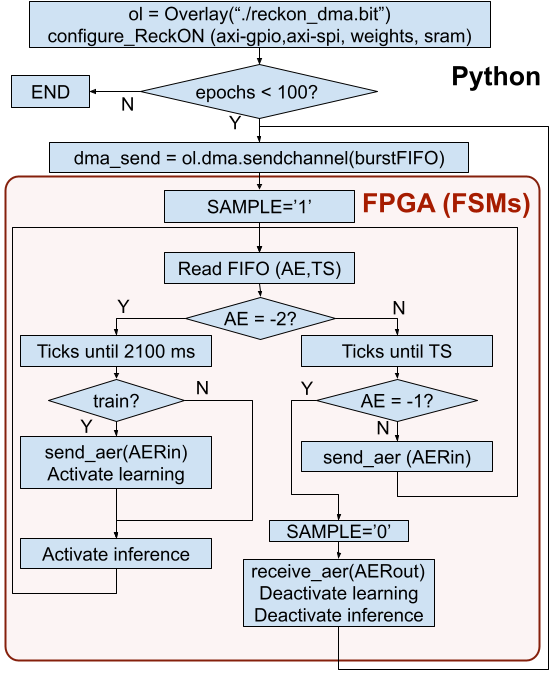}
 \caption{ReckOn test flow diagram}
  \label{fig:dflujo}
\end{figure}

We deploy the MPSoC-based ReckOn system following the flow diagram shown in Fig.~\ref{fig:dflujo}. 
A first configuration phase uses the SPI to initialize the RSNN parameters and state, the configuration registers, and the weight memories with random values. Then, the actual training phase takes place and unfolds over a given number of epochs (default: $100$). The FSM first triggers the $SAMPLE$ signal to ReckOn to indicate the start of a new sample, which itself contains a sequence of events described as (AE,TS) tuples (Section~\ref{sec:implementation}), where AE will be written in the address field of the input AER bus, and TS will determine the waiting time, in ticks, until the event is processed. Two special AE addresses: (i) $-2$ is used to indicate the classification label (per sample) or regression value (per timestep), which is used as a target during training and as a ground truth during inference to determine the obtained accuracy, and (ii) $-1$ is used to indicate the end of a sample. 
Once the end of the sample is reached, the SAMPLE signal is deasserted and the output of ReckOn is retrieved.

\begin{figure*}[h]
\centering
 \includegraphics[scale=0.265]{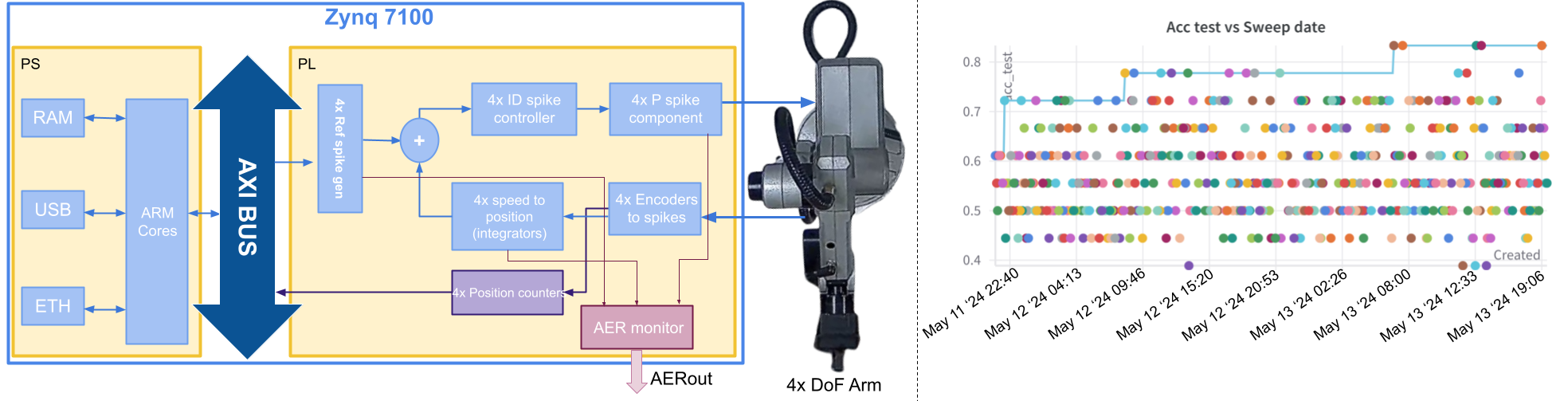}
 \caption{Left: ED-Scorbot SPID controllers and spiking activity recording scenario for the collected dataset. Right: Accuracy on tests during hyperparameter search.}
  \label{fig:ED_Scorbot_SPID}
\end{figure*}

\begin{figure}[t!]
\centering
 \includegraphics[scale=0.29]{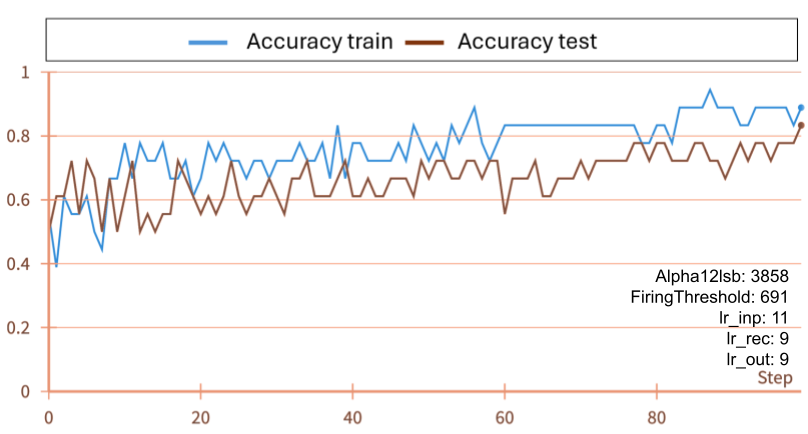}
 \caption{Accuracy in train/test for best hyperparameters.}
  \label{fig:ACC_Scorbot}
\end{figure}

The first experiment, implemented for validation, replicates the testbench from \cite{Frenkel_ReckOn_git} for the delayed cue accumulation scenario (also known as \textit{T maze}) of a mobile robot avoiding obstacles as in \cite{reckon_frenkel_2022}, preserving the accuracies of 100\% on the training set and 98\% on the test set.
To further demonstrate our MPSoC-based ReckOn system, we deploy it in an adaptive robotic arm control use case, where the goal is to determine if there is a weight attached to a robotic arm gripper while it is executing a lemniscate trajectory. The robotic arm gripper is the ED-Scorbot \cite{EDScorbot23}, which has four degrees of freedom (DoF). Its motor controllers are spike-based PIDs (SPID) implemented on a Zynq-7100 FPGA, which can receive target joint angles (as spike frequency references) every 100ms to execute any trajectory within its working area. Figure \ref{fig:ED_Scorbot_SPID} (left panel) shows a block diagram of the scenario. Each SPID (split into ID + P blocks) receives a spiking signal representing the error between the input spiking target angle and the current angle of the joint, also as a spiking signal. The spiking output of each SPID is used to drive the corresponding joint's motor. Each of these spiking signals has polarity.
These spiking activities have been recorded while the robot was executing 18 different lemniscate-shape trajectories~\cite{Lemniscate_EDScorbot_git}. Each trajectory was repeated with and without a 1-kg weight attached to the gripper. While maintaining the direction changes of the joints, the recorded spiking activity has been filtered and shrunk to a duration of $2250\,$ms, corresponding to the sample duration to be learned and classified by ReckOn. We have split the dataset into 50\% of the samples for training and the rest for testing.

We used a 24-200-2 network topology on ReckOn, as per Fig.~\ref{fig:diagramabloques} (right panel), with 200 recurrent neurons, 24 input neurons, corresponding to the 6 spiking activity sources from the 4 SPIDs of the dataset, and two output neurons denoting the two possible decisions, i.e.~ no weight attached to the gripper or 1-kg. In order to determine the best hyperparameters for the time constants, firing thresholds, and learning parameters of this RSNN, we have used the \textit{Weights \& Biases} tool \cite{wandb_web} to automate the hyperparameter search while ReckOn is running online in the MPSoC. Figure \ref{fig:ED_Scorbot_SPID} (right panel) shows the test accuracy evolution for 500 executions with different hyperparameters with 100 epochs/execution. The best hyperparameters lead to an accuracy of 88.9\% on the training set and of 83.3\% on the test set (Figure \ref{fig:ACC_Scorbot}). The measured peak rate of input events processed via the AER input bus of ReckOn amounts to 3.8M events per second, with an average of 18k events per second from Python experiments.

The resource utilization in the PL amounts to 36\% of the available LUTs (30\% for ReckOn), 47\% of the BRAM resources (30\% for ReckOn), and 12\% of DSP blocks.

\section{Conclusions}
In this work, the spiking neural network accelerator ReckOn, an open-source digital RSNN that embeds online learning on temporal tasks, has been deployed on a Xilinx MPSoC platform that is part of a Pynq-ZU board. The different techniques and circuits used for this implementation and their control from a Jupyter Notebook are presented. We deployed it for an adaptive robotic arm control use case, where the objective is to detect the weight on a robotic arm gripper, demonstrating the ability of the proposed system to learn complex temporal dependencies in the real world. 
The required resource utilization in the PL amounts to about 30\%. These are promising results for a system that is able to carry out training and inference for an RSNN without any reliance on cloud systems. This MPSoC system therefore directly contributes to low-power, low-latency neuromorphic Edge-AI applications.

\bibliographystyle{IEEEtran}
\bibliography{biblio}

\end{document}